\documentclass[review]{elsarticle}
\usepackage{lineno,hyperref}
\modulolinenumbers[5]
\journal{Journal of \LaTeX\ Templates}
\begin{document}
\begin{frontmatter}

\title{$\mathcal{PT}$-Supersymmetric Square Well and Barrier}

\author[mymainaddress]{T. Koohrokhi\corref{mycorrespondingauthor}}
\cortext[mycorrespondingauthor]{Corresponding author}
\ead{t.koohrokhi@gu.ac.ir}

\address[mymainaddress]{Department of Physics, Faculty of Sciences, Golestan University, Gorgan, Iran}

\begin{abstract}
The Parity-Time ($\mathcal{PT}$) symmetric potentials are derived by non-Hermitian supersymmetric quantum mechanics for square well and barrier. These $\mathcal{PT}$-supersymmetric square well and barrier The partners have complex partners. The partners are isospectral with real energies. $\mathcal{PT}$-symmetry is only unbroken for the bound states.\\

\end{abstract}


\end{frontmatter}

%\linenumbers

%
%****************************section 1 : Introduction *******************************************
%
\section{Introduction}
During the past few decades, the two new developments in quantum mechanics, namely $\mathcal{PT}$-symmetric quantum mechanics ($\mathcal{PT}$ QM) and supersymmetric quantum mechanics (SUSY QM), have been the subjects of intense theoretical and experimental researches [1,2]. These theory-based achievements have brought for the quantum systems some important fundamental concepts as well as powerful algebra tools, individually. The $\mathcal{PT}$ QM has indicated non-Hermitian Hamiltonians can have real energy spectra by replacing the mathematical condition of Hermiticity by the physical condition that Hamiltonian have an unbroken PT-symmetry.

On the other hand, SUSY QM based on factorization and operator methods has discovered new mechanism to denotes the shape invariant potentials (SIPs) are analytically solvable. Moreover, in the SUSY QM by introducing the superpotential and Hamiltonians hierarchy have been demonstrated partner potentials have the same energy spectra (except one level).

%A symmetry of the Hamiltonian (or Lagrangian) can be spontaneously broken if the lowest energy solution does not respect that symmetry, as for example in a ferromagnet, where rotational invariance of the Hamiltonian is broken by the ground state.

The combination of $\mathcal{PT}$ QM and SUSY QM have been created another new development quantum mechanics named $\mathcal{PT}$-supersymmetric quantum mechanics ($\mathcal{PT}$ SUSY QM). This is also approved by some theories and experiments.

 \section{Complexified Supersymmetric Square Well}
The solutions of time independent of Schr\"{o}dinger equation for the simplest one-dimensional problem in quantum mechanics, i.e. particle in a square well, is known as,
\begin{equation}
\left \{\begin{array}{ll} \psi_{c}(x)=A\sin(kx) \\ \psi_{t}(x)=B\cos(kx) \end{array}
\right.
\end{equation}
where $k^{2}=E$ is square of the wave number ($\hbar=2m=1$). In SUSY QM, superpotential is defined as the logarithmic derivative of wave function $W(x)=-\psi^{\prime}(x)/\psi(x)$. Therefore, superpotentials for the wave functions of the Eq.(1) are the cotangent (denotes by subscript $"c"$) and the tangent (denotes by subscript "t") functions. Now, we introduce  a new form of superpotentials by adding an imaginary functions linearly,
\begin{equation}
\left \{\begin{array}{ll} W_{c}(x)= -k\cot(\alpha x)+\textrm{i}f_{c}(x) \\
W_{t}(x)=k\tan(\alpha x)+\textrm{i}f_{t}(x)
\end{array}
\right.
\end{equation}
According the SUSY QM, the partner potentials are obtained by [],
\begin{equation}
V_{1,2}(x)=W^{2}(x)\mp W^{\prime}(x)
\end{equation}
The upper and lower signs are relate to $V_{1}$ and $V_{2}$, respectively. By putting the superpotentials $W_{c}(x)$ and $W_{t}(x)$ to Eq. (3), we get,
\begin{eqnarray}
V_{1,2}(x)&=&k(k\mp \alpha)\left \{\begin{array}{ll} \csc^{2}(\alpha x)\\ \sec^{2}(\alpha x)\end{array}
\right \}-k^{2}-\left \{\begin{array}{ll} f^{2}_{c}(x)\\f^{2}_{t}(x)\end{array}
\right \}\nonumber\\
&+&\textrm{i}\left \{\begin{array}{ll} \mp f^{\prime}_{c}(x) -2k\cot(\alpha x)f_{c}(x)\\ \mp f^{\prime}_{t}(x)+2k\tan(\alpha x)f_{t}(x)\end{array}
\right \}
\end{eqnarray}
If the $V_{1}$ and $V_{2}$ potentials are similar in shape and differ only in the parameters that appear in them, then they are said to be shape invariant. The remainder that is defined as $R_{1}=V_{2}(k,x)-V_{1}(k+\alpha ,x)$, equals with,
\begin{equation}
R_{1}= \alpha(\alpha+2k)+\textrm{i}2\left \{\begin{array}{ll} f^{\prime}_{c}(x)+\alpha\cot(\alpha x)f_{c}(x)\\ f^{\prime}_{t}(x)-\alpha\tan(\alpha x)f_{t}(x)\end{array}
\right \}
\end{equation}
Accordingly, partner potentials are shape invariant only if the bracket term of the remainder to be zero. As a result, the functions $f_{c}(x)$ and $f_{t}(x)$ are determined by this constraint as,
\begin{equation}
\left \{\begin{array}{ll} f_{c}(x)=q\csc(\alpha x)\\f_{t}(x)=q\sec(\alpha x) \end{array}
\right.
\end{equation}
where $q$ is an arbitrary constant. By setting the $\alpha=k$, the superpotentials are gained as,
\begin{equation}
\left \{\begin{array}{ll} W_{c}(x)=-k\cot(kx)+\textrm{i} q\csc(kx) \\ W_{t}(x)=k\tan(kx)+\textrm{i} q\sec(kx) \end{array}
\right.
\end{equation}
Proportionally, the complex superpartners are,
\begin{equation}
\left \{\begin{array}{ll} V_{1}(x)= -q^{2}\left \{\begin{array}{ll} \csc^{2}(kx)\\ \sec^{2}(kx)\end{array}
\right \}-k^{2}+\textrm{i}qk\left \{\begin{array}{ll} -\cot(kx)\csc(kx)\\ \tan(kx)\sec(kx)\end{array}
\right \}  \\
 V_{2}(x)= (2k^{2}-q^{2})\left \{\begin{array}{ll} \csc^{2}(kx)\\ \sec^{2}(kx)\end{array}
\right \}-k^{2}+\textrm{i}3qk\left \{\begin{array}{ll} -\cot(kx)\csc(kx)\\ \tan(kx)\sec(kx)\end{array}
\right \}
\end{array}
\right.
\end{equation}
In unbroken SUSY, the wave function is obtained by $\psi(x)=\exp(-\int W(x)dx)$. As a result we have,
\begin{equation}
\left \{\begin{array}{ll} \psi_{c}(x)=A\sin(kx)\exp\left\{\textrm{i} \frac{q}{k}\ln\left[\csc(kx)-\cot(kx)\right]\right\} \\ \psi_{t}(x)=B\cos(kx)\exp\left\{\textrm{i} \frac{q}{k}\ln\left[\sec(kx)+\tan(kx)\right]\right\} \end{array}
\right.
\end{equation}
Finally, the complete answer is gotten by,
\begin{equation}
\psi(x)=\psi_{c}(x)+\psi_{t}(x)
\end{equation}
The probability densities $|\psi(x)|^{2}$ are equal for both complex (Eq. (9)) and real wave functions (Eq. (1)).

%****************************** Section 3: Complex Square Well ************************************
 \section{Infinite Square Well}
Now consider an infinite square well in one dimension with length $L=\pi$. The boundary conditions,
\begin{equation}
\psi(0)=\psi(L)=0
\end{equation}
require that,
\begin{equation}
\left \{\begin{array}{ll} B=0 \\ k_{n}=n+1 \end{array}
\right.
\end{equation}
where $n=0,1,2,...$ . The remainder is,
\begin{equation}
R_{1}=3k^{2}_{n}
\end{equation}
and according the unbroken SUSY, $E_{0}=0$, the energy spectrum is,
\begin{equation}
E_{n}=k^{2}_{n}-1=n(n+2)
\end{equation}
The potentials $V_{1}=-1$ ($q=0$ and $k=1$) and real and imaginary parts $V_{1cr}$ ($V_{1tr}$) and $V_{1ci}$ ( $V_{1ti}$), respectively, for typically values $q=2$ and $k=1$ are depicted in Fig. (1). The Fig. (2) illustrates the potentials $V_{2c}$ ($V_{2t}$) ($q=0$ and $k=1$) and real and imaginary parts $V_{2cr}$ ($V_{2tr}$) and $V_{2ci}$ ( $V_{2ti}$), respectively, for typically values $q=2$ and $k=1$.

In SUSY QM, superpotential is defined as the logarithmic derivative of wave function $W(x)=-\psi^{\prime}(x)/\psi(x)$ (for simplicity $\hbar=2m=1$). Now, assume a plane wave $\exp(\textrm{i}kx)$, that the new form its superpotential is made by adding a function $f(x)$ as,
\begin{equation}
W(x)=-\textrm{i}k+f(x)
\end{equation}
where $k^{2}=E$ is the wave number and $E$ is the energy. According the SUSY QM, the partner potentials are obtained by [],
\begin{equation}
V_{1,2}(x)=W^{2}(x)\mp W^{\prime}(x)
\end{equation}
The upper and lower signs are relate to $V_{1}$ and $V_{2}$, respectively. Therefor, for the superpotential $W(x)$ Eq. (1), we have,
\begin{equation}
V_{1,2}(x)=-k^{2}+f^{2}(x)+\textrm{i}2kf(x)\mp f^{\prime}(x)
\end{equation}
If the $V_{1}$ and $V_{2}$ potentials are similar in shape and differ only in the parameters that appear in them, then they are said to be shape invariant. The remainder is defined as,
\begin{eqnarray}
R_{1}&=&V_{2}(k,x)-V_{1}(k+\alpha ,x)\nonumber\\
&=&2k\alpha+\alpha^{2}+2\left[\textrm{i}\alpha f(x)+ f^{\prime}(x)\right]
\end{eqnarray}
Accordingly, partners are shape invariant if the bracket term of the remainder to be zero. As a result, the function $f(x)$ is obtained by this constraint,
\begin{equation}
f(x)=q\exp(-\textrm{i}\alpha x)
\end{equation}
where $q$ is an arbitrary constant. Therefore, the superpotential gain if we put $\alpha=k$,
\begin{equation}
W^{R}(x)=-\textrm{i}k+q\exp(-\textrm{i}kx)
\end{equation}
the superscript $R$ indicates the wave moving to the right. Proportionally, the complex superpartners are,
\begin{equation}
V^{R}_{1,2}(x)=-k^{2}+q^{2}\exp(-\textrm{i}2k x)+\textrm{i}qk(\pm 1-2)\exp(-\textrm{i}k x)
\end{equation}
and the wave function is,
\begin{equation}
\psi^{R}(x)=N\exp\left\{-\int W(x)dx\right\}=N\exp(\textrm{i}k x)\exp\left\{\frac{-\textrm{i}q\exp(-\textrm{i}k x)}{k}\right\}
\end{equation}
The information of the wave moving to the left obtain if we replace $k$ by $-k$. The superpotential is,
\begin{equation}
W^{L}(x)=\textrm{i}k+q\exp(\textrm{i}kx)
\end{equation}
and the complex superpartners are,
\begin{equation}
V^{L}_{1,2}(x)=-k^{2}+q^{2}\exp(\textrm{i}2k x)-\textrm{i}qk(\pm 1-2)\exp(\textrm{i}k x)
\end{equation}
and the wave function is,
\begin{equation}
\psi^{L}(x)=N\exp\left\{-\int W(x)dx\right\}=N\exp(-\textrm{i}k x)\exp\left\{\frac{\textrm{i}q\exp(\textrm{i}k x)}{k}\right\}
\end{equation}
Therefore, complete answer is,
\begin{equation}
\psi(x)=A\psi^{R}(x)+B\psi^{L}(x)
\end{equation}

%*************************fig. 1*************************
\begin{figure}
  \includegraphics[width=0.95\textwidth]{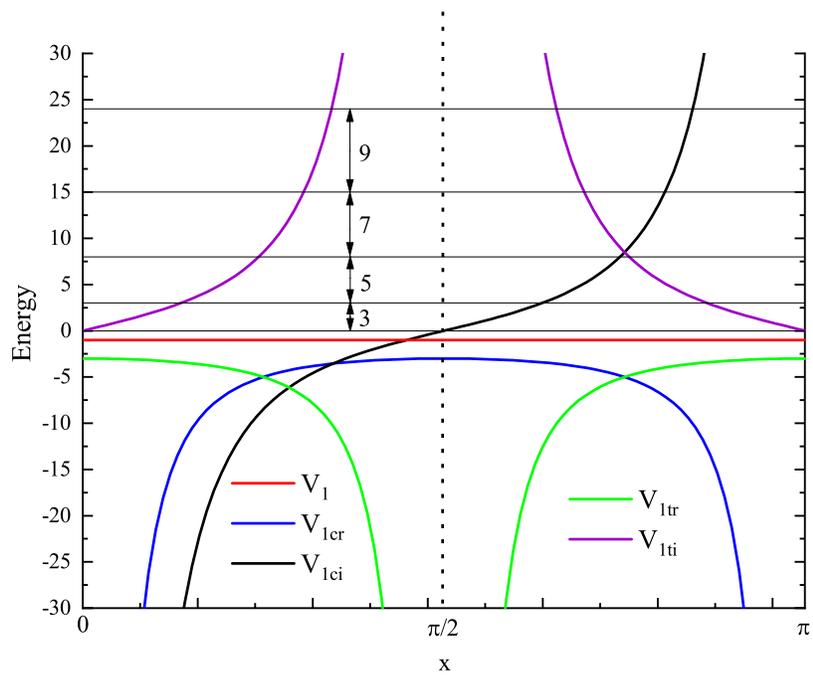}
\caption{The potentials $V_{1}=-1$ ($q=0$ and $k=1$) and real and imaginary parts $V_{1cr}$ ($V_{1tr}$) and $V_{1ci}$ ( $V_{1ti}$), respectively, for typically values $q=2$ and $k=1$} \label{fig:1}
\end{figure}
\clearpage
%*************************fig. 2*************************
\begin{figure}
  \includegraphics[width=0.95\textwidth]{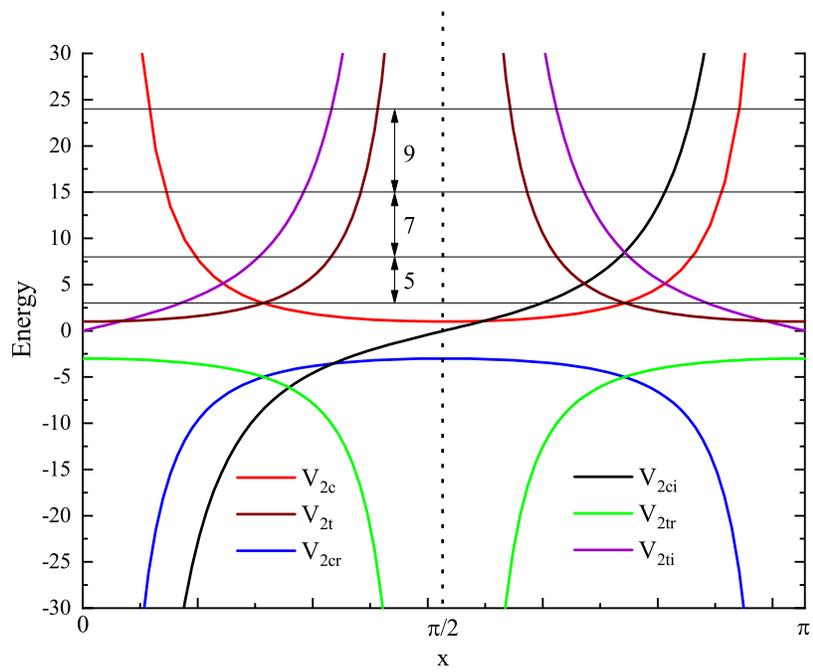}
\caption{The potentials $V_{2c}$ ($V_{2t}$) ($q=0$ and $k=1$) and real and imaginary parts $V_{2cr}$ ($V_{2tr}$) and $V_{2ci}$ ( $V_{2ti}$), respectively.} \label{fig:2}
\end{figure}
\clearpage

%******************************references***************************

\end{document}